\documentclass{aa}
\usepackage{graphics}
\usepackage{epsfig}
\def\E$\gamma${E_$\gamma$}

\def \deg      {$^{\circ}$}

\begin{document}

%\thesaurus{
%               03                    % Main Journal (?)
%              (13.07.2;              % Gamma rays: observations
%               11.01.2;              % Galaxies: active
%               11.17.4 3C~273);      % Galaxies: quasars: individual: 3C273
%              }
%

\title{INTEGRAL and XMM-Newton observations towards the unidentified MeV source GRO~J1411-64}

\author{Diego F. Torres$^{1}$,
Shu Zhang$^{2}$, Olaf Reimer$^{3}$, Xavier Barcons$^{4}$, Amalia
Corral$^{4}$, Valent\'{\i} Bosch-Ramon$^{5}$, Josep M.
Paredes$^{5}$, Gustavo E. Romero$^{6}$, Jin Lu Qu$^{2}$, Werner
Collmar$^{7}$, V. Sch\"onfelder$^{7}$ \& Yousaf Butt$^{8}$ }

\institute{Institut de Ci\`encies de l'Espai (IEEC/CSIC) , Campus
UAB, Facultat de Ci\`encies, Torre C5-parell, 2$^{\rm da}$ planta,
08193 Barcelona, Spain \and  Laboratory for Particle Astrophysics,
Institute of High Energy Physics, Beijing 100049, China \and W. W.
Hansen Experimental Physics Laboratory, Stanford University,
Stanford, CA 94305, USA \and Instituto de F\'{\i}sica de Cantabria
(CSIC-UC), E-39005 Santander, Spain \and Universitat de Barcelona,
Av. Diagonal 647, E08028, Barcelona, Spain \and Instituto
Argentino de Radioastronomia, CC5, 1894, Villa Elisa, Argentina
\and Max-Planck-Institut f\"ur extraterrestrische Physik, PO Box
1603, D-85740 Garching, Germany \and Harvard-Smithsonian Center
for Astrophysics, 60 Garden St., Cambridge, MA 02138, USA  }

\offprints{Diego F. Torres} \mail{dtorres@ieec.uab.es}

\date{Received  / Accepted }

\titlerunning{GRO~J1411-64}
\authorrunning{D. F. Torres et al.}

%context aims methods results

% context heading (optional)
\abstract {The COMPTEL unidentified source GRO~J1411-64 was
observed by INTEGRAL, and its central part, also by XMM-Newton.}
{The aim of these observations and subsequent analysis and
theoretical investigations pursued is to shed light upon the
nature of such source.} {Usual data analysis techniques of
INTEGRAL and  XMM-Newton were used in this work.} {The data
analysis shows no hint for new detections at hard X-rays. The
upper limits in flux herein presented constrain the energy
spectrum of whatever was producing GRO~J1411-64, imposing, in the
framework of earlier COMPTEL observations, the existence of a peak
in power output located somewhere between 300-700 keV for the
so-called low state. The Circinus Galaxy is the only source
detected within the 4$\sigma$ location error of GRO~J1411-64, but
can be safely excluded as the possible counterpart:  the
extrapolation of the energy spectrum is well below the one for
GRO~J1411-64 at MeV energies. 22 reliable and statistically
significant sources (likelihood $> 10$) were extracted and
analyzed from XMM-Newton data. Only one of these sources, XMMU
J141255.6-635932, is spectrally compatible with GRO~J1411-64
although the fact the soft X-ray observations do not cover the
full extent of the COMPTEL source position uncertainty make an
association hard to quantify and thus risky. }{ The unique peak of
the power output at high energies (hard X-rays and gamma-rays)
resembles that found in the SED seen in blazars or microquasars,
and might suggest that a similar scenario is at work. However, an
analysis using a microquasar model consisting on a magnetized
conical jet filled with relativistic electrons which radiate
through synchrotron and inverse Compton scattering with star,
disk, corona and synchrotron photons shows that it is hard to
comply with all observational constrains. This fact and the
non-detection at hard X-rays introduce an a-posteriori question
mark upon the physical reality of this source, which is discussed
in some detail. }
% aims heading (mandatory)

\keywords{X-rays: stars -- $\gamma$-rays: observations}

\maketitle

\section{Introduction}

The COMPTEL experiment onboard the Compton Gamma-Ray Observatory
(CGRO) surveyed the sky in the $0.75-30$ MeV energy range and
detected about a dozen sources of Galactic origin. Among them,
pulsars, stellar black-hole candidates, and supernova remnants
were identified by the telescope (Sch\"onfelder et al. 2000). In
addition, nine unidentified sources were also reported in the
first COMPTEL Catalog. Zhang et al. (2002) have reported the
discovery of a new source belonging to this group. It is a
variable source located near the Galactic plane.
The fluxes of this COMPTEL unidentified source along the first
five phases of the Compton satellite were derived by  Zhang et al.
(2002), and are shown, together with the corresponding confidence
level of each detection, in Table 1. Time periods and effective
exposures for different observations are also listed there. The
best detection significance, at more than 7$\sigma$, was obtained
in the 1-3 MeV band by combining 7 viewing periods (VPs, the
periods of observations in the Compton satellite) during 1995
March-July, VP 414-424. The probability of randomly detecting a
source with such high level of significance in the COMPTEL 1-3~MeV
band was estimated to be 5.4$\times$10$^{-10}$; obtained by taking
into account the trials for searching in all individual VPs in
four energy bands from the beginning of the Compton mission in
1991 to its second reboost in 1997 (Zhang et al. 2002). The most
likely source location, i.e. the maximum of the likelihood ratio
distribution in the 1-3 MeV band for the combination of VPs in
1995,  is  $(l,b) = (311.5$\deg ,$-2.5$\deg$)$. Therefore, the
source was referred to as GRO J1411-64. The flare flux (where
following the report by Zhang et al. the detection significance of
7.2$\sigma$ was measured, see the discussion below) was about 0.3
Crab, which would make of GRO J1411-64 one of the strongest
sources at low galactic latitudes, second only to Crab itself. The
{\it flare duration} was several months, the longest one compared
to all other $\gamma$-ray sources that were monitored almost
continually by COMPTEL. The rather steep spectral shape obtained
while the source was flaring would predict a bright, hard X-ray
source, if there is no break in the spectrum, which is explored
here.

\begin{table*}[tbh]
\begin{center}
\caption{The GRO~J1411-64 observations: effective exposures,
fluxes including detection significance ($\chi_1^2$) or 2$\sigma$
flux upper limits for the observations reported by Zhang et al.}
\begin{tabular}{ccccccc}\hline
\multicolumn{1}{c}{TJD}&\multicolumn{1}{c}{Effective exposure }&\multicolumn{4}{c}{Flux (10$^{-5}$ ph cm$^{-2}$ s$^{-1}$)}\\
\multicolumn{1}{l}{ }&\multicolumn{1}{c}{days}&\multicolumn{1}{c}{0.75-1
MeV}&\multicolumn{1}{c}{1-3  MeV}&\multicolumn{1}{c}{3-10
MeV}&\multicolumn{1}{c}{10-30 MeV}\\  \hline
8392-8943 (Phase 1)     &9.29   &6.0$\pm$3.1 (2.3$\sigma$)  &$<$6.0   &1.7$\pm$1.6 (1.1$\sigma$)  &0.8$\pm$0.6 (1.1$\sigma$)\\
8943-9216 (Phase 2)     &5.87   &4.2$\pm$3.5 (1.3$\sigma$)  &$<$10.1   &$<$3.9   &1.8$\pm$0.9 (2.0$\sigma$)\\
9216-9629 (Phase 3)     &7.45   & $<$12.5   &6.7$\pm$3.8 (1.8$\sigma$)  &2.2$\pm$1.6 (1.9$\sigma$)  &$<$1.6 \\
9629-9993 (Phase 4)     &8.42   &13.7$\pm$3.8 (4.0$\sigma$)   &21.2$\pm$3.5 (5.9$\sigma$)  &4.6$\pm$1.6 (2.9$\sigma$)  &$<$1.2 \\
9993-10371(Phase 5)     &4.23   &18.6$\pm$4.7 (4.4$\sigma$)  &$<$14.7   &$<$4.8   &$<$2.6 \\
9797-9923 (VPs 414-424) &5.44   &19.9$\pm$4.6 (4.6$\sigma$)  &36.6$\pm$4.7 (7.7$\sigma$)  &5.4$\pm$2.3 (2.1$\sigma$)  &$<$1.8 (2.0$\sigma$)\\
8392-10371(Phases 1-5) &35.26  &8.4$\pm$1.7 (5.1$\sigma$)  &6.1$\pm$1.6 (4.0$\sigma$)
&1.6$\pm$0.8 (2.4$\sigma$)  &0.5$\pm$0.3 (2.0$\sigma$)\\  \hline
\end{tabular}\end{center}
\label{tab:flux}
\end{table*}

That the source was variable within timescales of several months
could be shown by two different methods by Zhang et al. (2002). On
one hand, a direct statistical approach shows that if the fluxes
of the individual viewing periods  of the first five Phases are
fitted with a constant flux, a $\chi^{2}$ value of 94.6 is derived
(1-3~MeV band). That value corresponds to a significance of
4.1$\sigma$ for a variable source; alternatively, to a probability
of $\sim$4.8$\times$10$^{-5}$ for a constant flux. On the other
hand, the variability index $I$ (Torres et al. 2001), used to
normalize the measured variability of this source to the measured
variability of other COMPTEL sources considered to be non-variable
(so as to encompass systematics), gives a value of 5.2. This is
above the 3$\sigma$-level and also classifies the source as likely
variable. However, it is unclear whether this second method would
apply to COMPTEL sources, due to the larger spread of indices for
detected COMPTEL pulsars.
In what follows, we present the results of the INTEGRAL
observations of this source, as well as of XMM-Newton observation
of its best location. We also  comment on what can be inferred out
of these observations concerning models of emission based on
microquasars.

%######################################################################
\section{Observation, data analysis, and results}
%######################################################################

GRO~J1411-64 was observed by INTEGRAL  during 2004 December 30 -
2005 January 6. In total, 102 science windows (scws) were carried
out to have 210 ks of effective exposure. Data were taken in
revolutions 270-272 using an hexagonal dithering mode. Due to the
relatively small field of view (FOV) of JEMX, for most of the time
the source was actually outside the scope of this instrument, and
the total effective exposure in JEMX was then lowered down to
roughly 83 ks.
%
%For JEMX, due to the relatively small FOV, for many time the source is
%close to the edge of the FOV. Although the total observational
%exposure of JEMX is also around 210 ks, but the effective exposure
%goes down to ~83 ks.
%
Data reduction was performed using the version 5.0 of the standard
Offline Science Analysis (OSA) software, and the spectra were
fitted with XSPEC of FTOOLS 5.3.1.

\begin{table*}[t]
\begin{center}
\caption{Detected INTEGRAL sources and their confidence level
($\sigma$), flux and flux error (in units of ct/s), in the
different energy ranges explored (in keV). } \vspace{0.2cm}
\begin{tabular}{llllllllll}
\hline
\multicolumn{1}{c}{Source Name}&\multicolumn{3}{c}{20-40 keV}&\multicolumn{3}{c}{40-80 keV}&\multicolumn{3}{c}{80-120 keV}\\
\multicolumn{1}{c}{ }&\multicolumn{1}{c}{$\sigma$} &\multicolumn{1}{c}{flux}&\multicolumn{1}{c}{flux error}&\multicolumn{1}{c}{$\sigma$}&\multicolumn{1}{c}{flux}&\multicolumn{1}{c}{flux error}&\multicolumn{1}{c}{$\sigma$}&\multicolumn{1}{c}{flux}&\multicolumn{1}{c}{flux error}\\  \hline

\hline
 GX 301-2                & 149&  10.72    &0.07   &12     &0.82  &0.07   &\ldots & \ldots & \ldots     \\
 Circinus Galaxy         &  38&  1.35     & 0.03   &20     &0.75  &0.04   &\ldots & \ldots &  \ldots    \\
 4U 1626-67              &  20&  1.95    &0.10   &\ldots &\ldots     & \ldots      &\ldots & \ldots  & \ldots     \\
 PSR B1509-58            &  16&   0.82  & 0.05  &13     &0.66       &0.05         &  8               &0.28 &0.04   \\
 H 1538-522              &  12&  2.06     & 0.16   &\ldots &\ldots     &\ldots       &\ldots         &\ldots &\ldots  \\
 1E 1145.1-6141          &  10&  1.77    & 0.18   &\ldots &\ldots     &\ldots       &  \ldots        &\ldots &\ldots   \\
 Cir X-1                 &  9 &  0.58    &0.06   &\ldots &\ldots     &\ldots       & \ldots        &\ldots &\ldots    \\
 NGC 4945                &  7 &   1.31   &0.19    &\ldots &\ldots     &\ldots       &  \ldots        &\ldots &\ldots   \\
 2RXP J130159.5-63580    &  7 &   0.32  &0.05   &\ldots &\ldots     &\ldots       &   \ldots        &\ldots &\ldots  \\
 4U 1323-62              &  34&   1.36   & 0.04  &17     &0.73  &0.04   & \ldots  &\ldots &\ldots   \\

\hline \hline
\end{tabular}
\end{center}
\end{table*}

\subsection{Images}

%%%%% Fig 1 %%%%%%%%%%%%%%%%%%%%%%%%%%%%%%%%%%%%%%%%%%%%
\begin{figure*}[t]
\centering
\includegraphics[clip=true,scale=0.4]{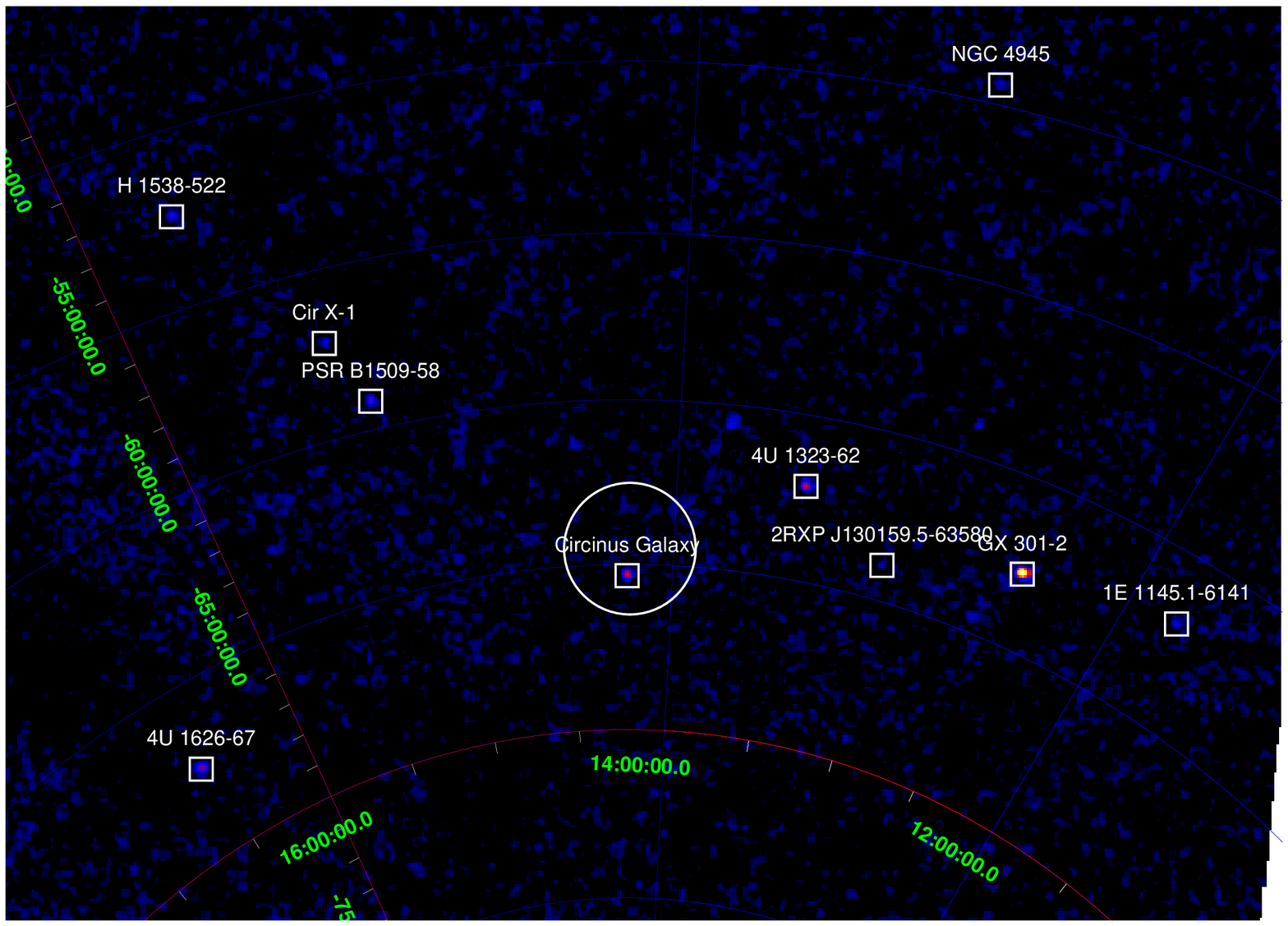}
\includegraphics[clip=true,scale=.4]{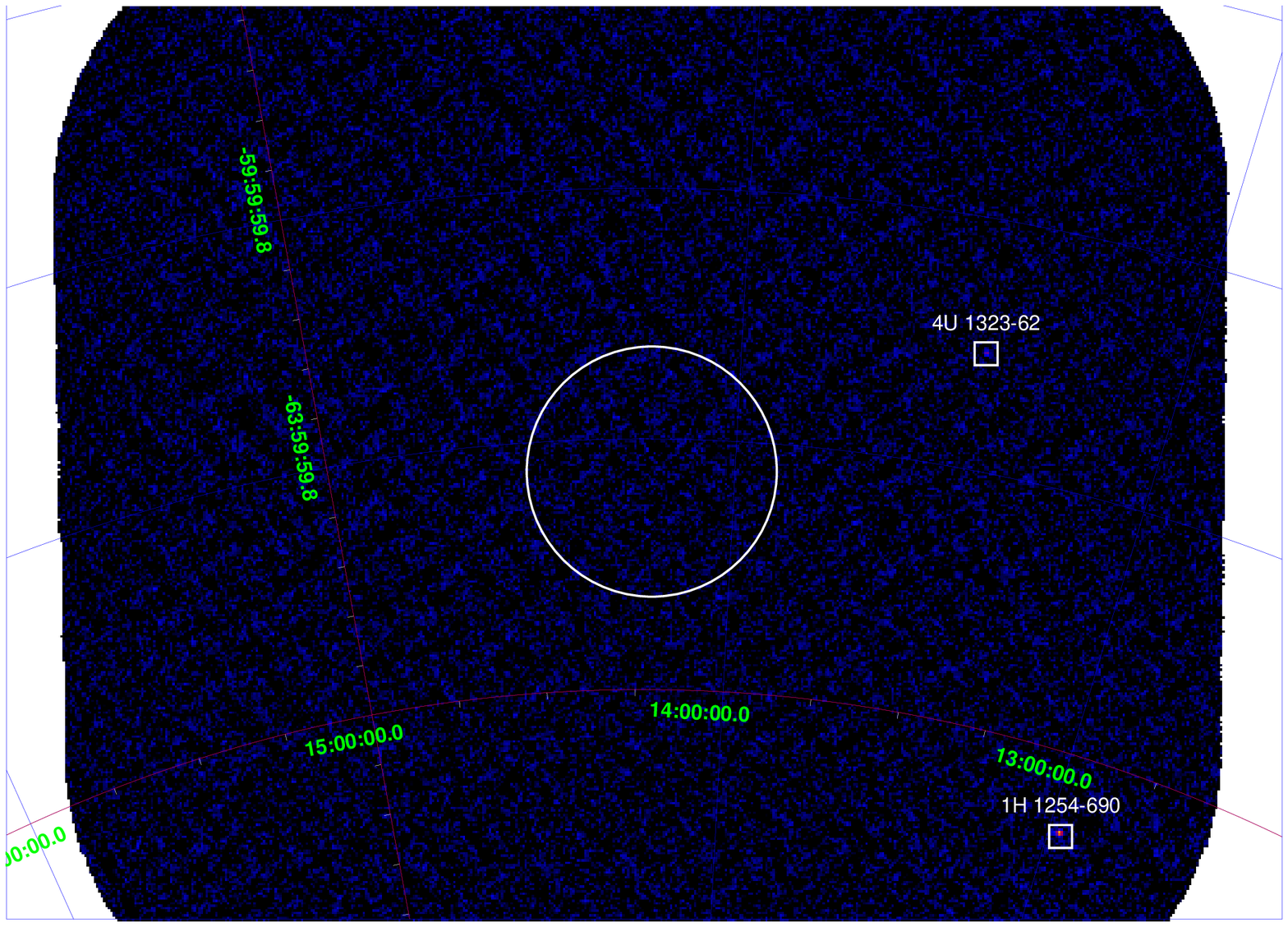}
\caption{Left: Sky map of the GRO~J1411-64 region as seen by
IBIS/ISGRI in the 20-40 keV range, by combining all data  obtained
in the observations performed during 2004 December 30 to 2005
January 6. See Table 2 for details of each detected source. Right:
Sky map of the GRO~J1411-64 region as seen by JEMX in 1.6-10 keV
range, by combining all data obtained  in the observations
performed during 2004 December 30 to 2005 January 6. The detected
two sources are 1H 1254-690 (122$\sigma$), and 4U 1323-62 (39
$\sigma$).}
\end{figure*}
%%%%% Fig 1 %%%%%%%%%%%%%%%%%%%%%%%%%%%%%%%%%%%%%%%%%%%%

No hint of signal was found for new hard X-ray sources within the
location uncertainty of GRO~J1411-64 from individual scws of the
INTEGRAL instruments. To improve the statistics,  mosaic maps were
obtained for IBIS/ISGRI and JEMX by combining all data. The images
of IBIS/ISGRI were produced in the energies 20-100 keV, see Figure
1 (left panel) for the map in the 20-40 keV band as an example.
The circle shows the 4$\sigma$ of error location of GRO~J1411-64
obtained by COMPTEL during its flare in 1995 (Zhang et al. 2002).
This circle is centered at the location of the MeV source, with a
radius of 2 degrees. From the possible counterparts of
GRO~J1411-64 discussed in Zhang et al.'s paper (2002), only the
Circinus Galaxy shows up in this error region as seen by INTEGRAL.
The most significant detection of Circinus Galaxy is detected at
20-40 keV, at a confidence level of 38$\sigma$. In the mosaic map,
there are 10 additional sources detected, among them the strongest
one is HMXB GX301-2, with 149$\sigma$ at 20-40 keV.

The mosaic map of JEMX is shown in Figure 1 (right panel) at
1.6-10 keV. Again the circle represents the 4$\sigma$  error
contour of COMPTEL GRO~J1411-64. No significant source feature is
visible from within this error region. The Circinus Galaxy was
most of the time outside the FOV of JEMX, and therefore it is not
visible due to the relatively small exposure. Only two sources are
detected from the combined JEMX data.  The strongest source is
1H1254-690, detected with a confidence level of 122$\sigma$ in the
3-10 keV range. For SPI, only a few sources are detected at
energies below 100 keV and the maps are empty at energies above
100 keV.  The Circinus Galaxy is at the 6$\sigma$ level in the
20-40 keV range, and it is the only source detected within the
location of GRO~J1411-64, the region of our search.

For the sources detected in IBIS/ISGRI and JEMX, only  GX 301-2,
Circinus Galaxy, PSR B1509-58, 4U 1323-62, and 1H1254-690 are of
interest for further investigation either due to their relatively
high detection significance (like 1H1254-690 and GX 301-2) or the
visible source signal extended in a  relatively wide range of
energies (like PSR B1509-58 and 4U 1323-62). GX 301-2 was already
reported by Kreykenbohm et al. (2004), from  early INTEGRAL data
corresponding to 24 scws. Only one scw was at pre-periastron, with
a flare detected to have its flux enhanced by a factor of 10 than
the low emission state of the orbital phase. The Circinus
Galaxy was also reported by Soldi (2005). %with 589 ks of INTEGRAL exposure.
%They found the source has rather high absorption.
For PSR B1509-58, an additional INTEGRAL observation of 1000 ks
exposure was carried out in 2005 and the data will be analyzed and
reported elsewhere, we thus refrain of further comment on this
source. 4U 1323-62 and 1H 1254-690 have to our knowledge not being
reported earlier. Hence the preliminary analysis carried out in
this paper sheds the first light on their properties in hard
X-rays.
%The detailed analysis on
%these two sources are in process and the results will be presented
%separately in another paper.

\subsection{Light curves}

%%%%% Fig 2 %%%%%%%%%%%%%%%%%%%%%%%%%%%%%%%%%%%%%%%%%%%%
\begin{figure*} [t]
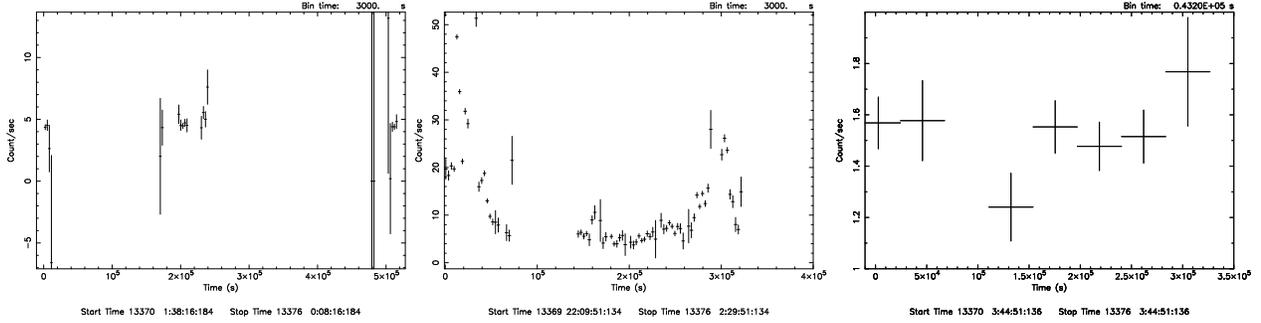

\centering
\includegraphics[clip=true,scale=.23, angle=-90]{1H1254_lc_3-10-bin3000.ps}
\includegraphics[clip=true,scale=.23, angle=-90]{GX301.lc_20-40-bin3000.ps}
\includegraphics[clip=true,scale=.23, angle=-90]{Circinus.lc_20-40-halfday.ps}
\caption{Left: Light curve of 1H1254-690 with each bin averaged
over 3000 seconds, observed by JEMX in the 1.6-10 keV range.
Middle: Light curve of GX 310-2 with each bin averaged over 3000
seconds, observed by IBIS/ISGRI in the 20-40 keV range. Right:
Light curve of the Circinus Galaxy with each bin averaged over
half day, observed by IBIS/ISGRI  in the 20-40 keV range.}
\end{figure*}
%%%%% Fig 2 %%%%%%%%%%%%%%%%%%%%%%%%%%%%%%%%%%%%%%%%%%%%

The light curve for the most significant JEMX source 1H1254-690 is
shown in Figure 2 (left panel), for the IBIS/ISGRI source GX301-2
in the middle panel, and for the Circinus Galaxy in the right
panel. The latter was considered a priori to be the a potential
counterpart of GRO~J1411-64, and was detected mainly by
IBIS/ISGRI. As shown in these light curves, hard X-ray emission
from the Circinus Galaxy and 1H1254-690 is rather constant,
whereas that from GX 301-2 is highly variable. GX 301-2 is an
eccentric hard X-ray binary system, famous for its rapid
outbursts, strongly correlated to its orbital phase. The burst
structure in this case is typical for GX 301-2, generally formed
during the time period of orbital phase pre- and post periastron
(Koh et al. 1997, Kreykenbohm et al. 2004, La Barbera et al.
2005). This source is, nonetheless, far from the region of
GRO~J1411-64, and thus it may not be related to it.

\subsection{Energy spectra}

%%%%% Fig 3 %%%%%%%%%%%%%%%%%%%%%%%%%%%%%%%%%%%%%%%%%%%%
\begin{figure*}
\centering
\includegraphics[clip=true,scale=.23, angle=-90]{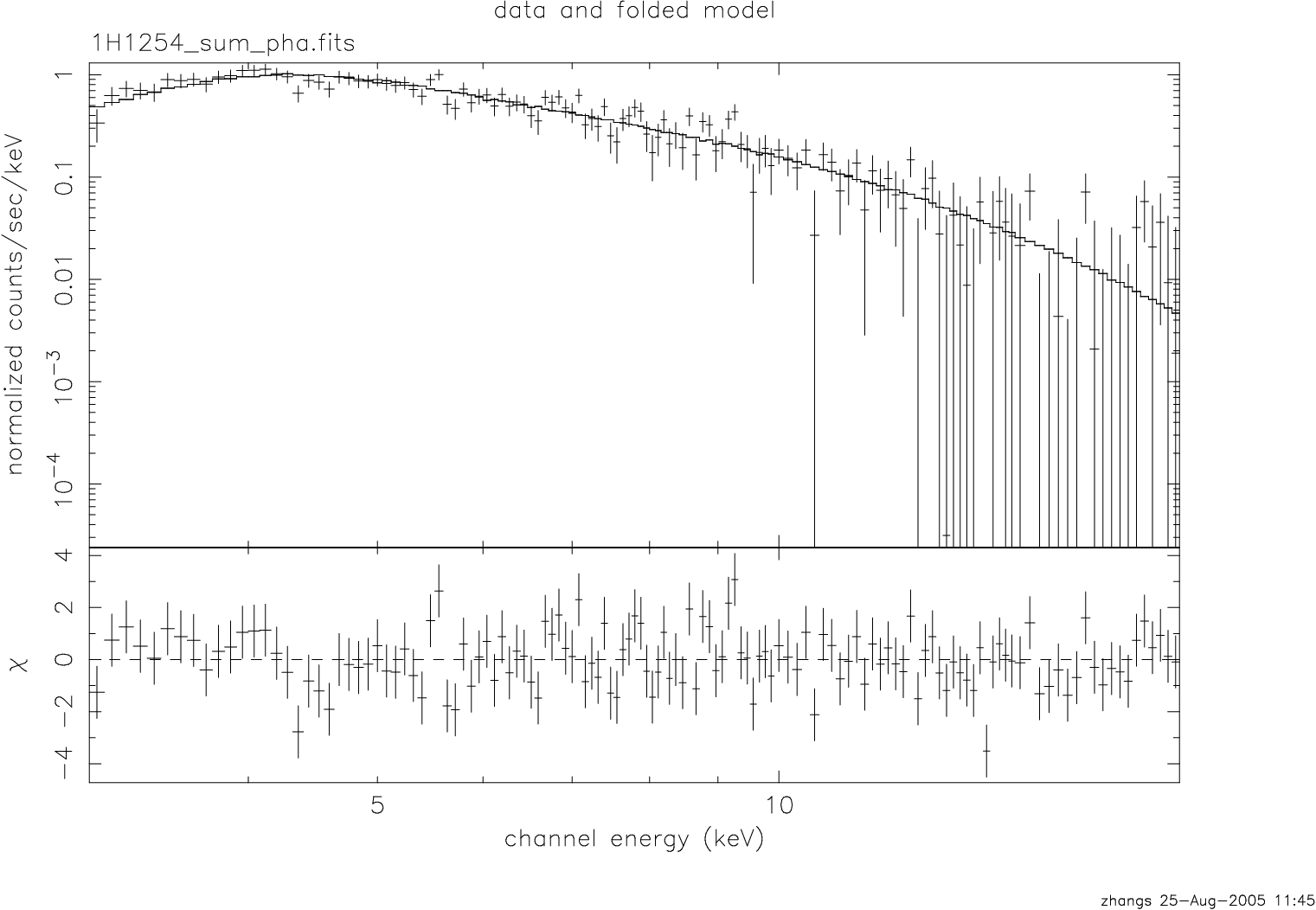}
\includegraphics[clip=true,scale=.23, angle=-90]{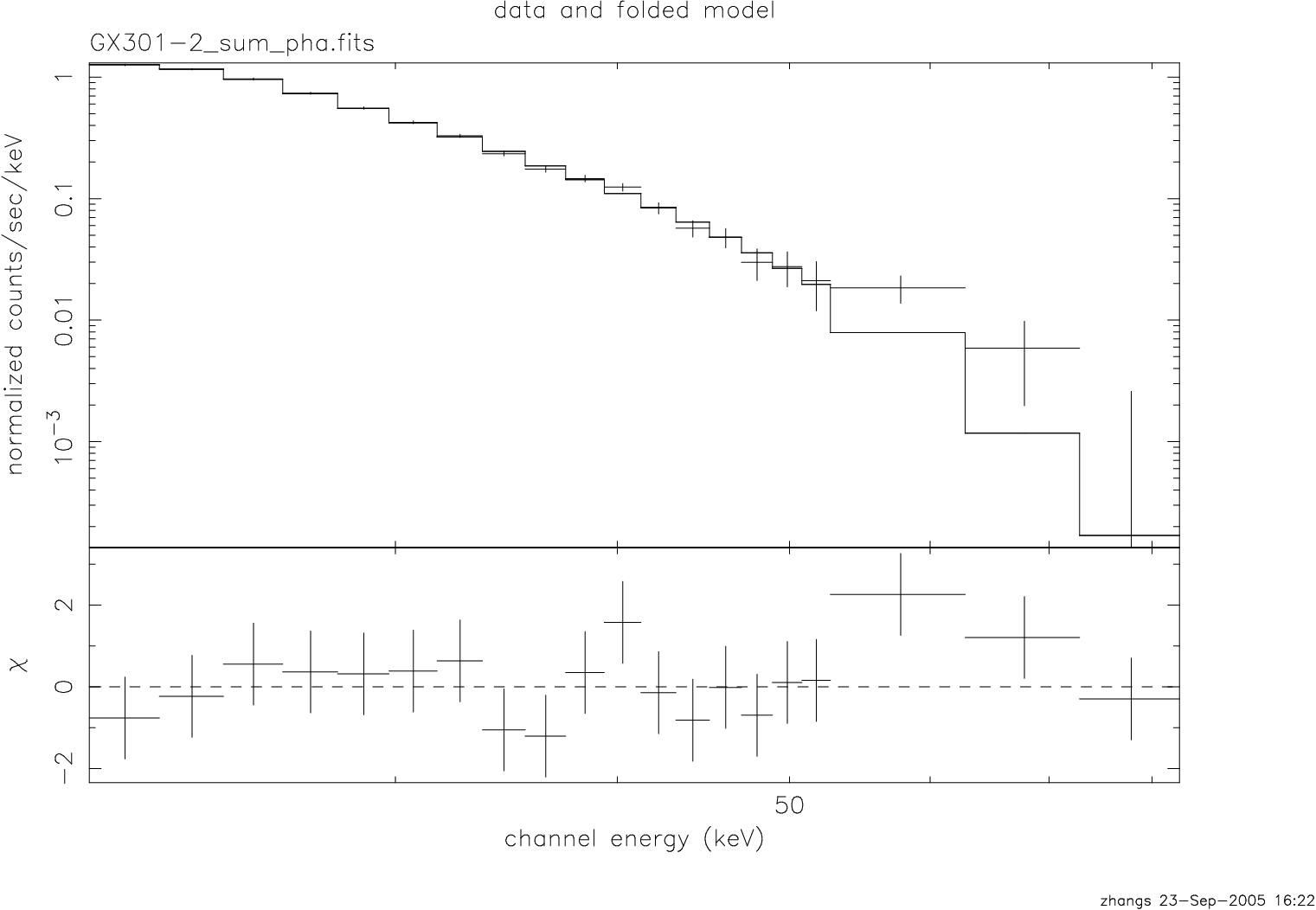}
\includegraphics[clip=true,scale=.23, angle=-90]{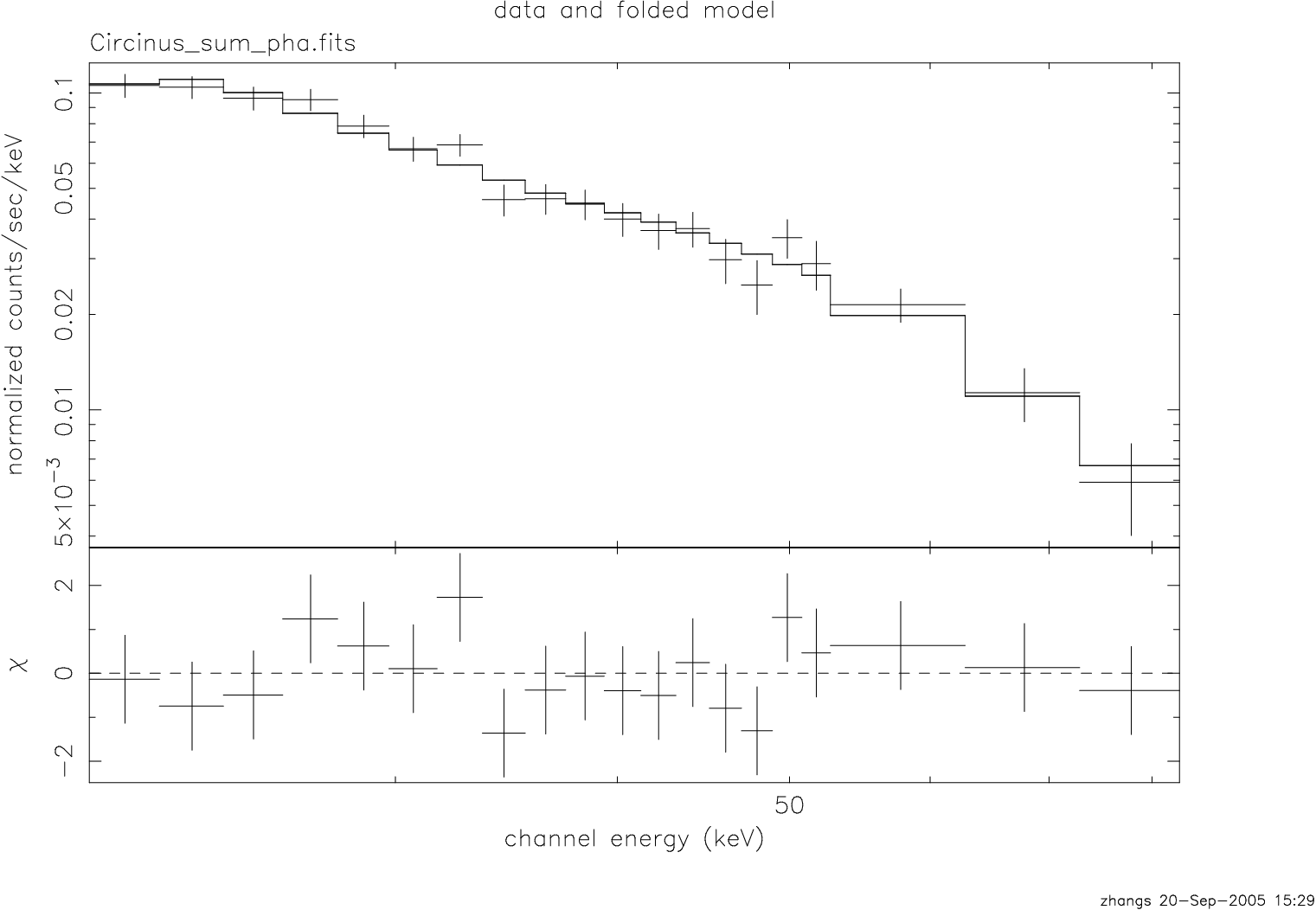}
\caption{Left: Spectrum of 1H1254-690 from the JEMX combined data.
The fit model is cutoffpl in XSPEC. Middle: Spectrum of GX 301-2
from IBSI/ISGRI combined data.  The fit model is cutoffpl in
XSPEC. Right: Spectrum of the Circinus Galaxy from IBIS/ISGRI
combined data. The fit model is cutoffpl plus wabs in XSPEC.}
\end{figure*}
%%%%% Fig 3 %%%%%%%%%%%%%%%%%%%%%%%%%%%%%%%%%%%%%%%%%%%%

Since the sources are mainly detected by the individual
instruments of INTEGRAL, the energy spectra of 1H1254-690, GX301-2
and the Circinus Galaxy, were analyzed with the observational data
from the corresponding instruments that detected them. The data
were summed up at the well calibrated energies of 3-20 keV for
JEMX, and photons with energies $\ge$ 20 keV detected by
IBIS/ISGRI were used in spectral fitting. 1H1254-690 is well
fitted by the cutoffpl model (see Eq.1) in XSPEC,
\begin{equation}
 f(E) = KE^{-\alpha} \exp{(-E/E_c)}
\end{equation}
where $K$ is a normalization factor (keV$^{-1}$cm$^{-2}$s$^{-1}$)
at 1 keV, $\alpha$ is the power law photon index and $E_c$ is the
energy of the exponential rolloff (keV). This model produces a
reduced $\chi^2$ of 1.2 for 133 dof (see Figure 3). The cutoffpl
model represents well the spectral data for GX 301-2 as well (see
Figure 3, middle panel), with a reduced $\chi^2$ of 0.9 (17 dof).
The former reported  cyclotron resonance scattering feature (CRSF)
at either 45 keV or 35 keV (La Barbera et al. 2005; Kreykenbohm et
al. 2004) are not visible in ISGRI, most probably due to the
relatively poor statistics of the data. The Circinus Galaxy was
investigated by Soldi et al. (2005);
%with a larger exposure of INTEGRAL (598 ks from 1998-2001).
models of cutoffpl plus wabs in XSPEC can fit the data well, with
a reduced $\chi^2$ of 1.1 (7 dof). The resulting parameters from
Soldi et al. were referred to in our fitting of the IBIS/ISGRI
observational data as follows: $N_H$ and $E_c$ were fixed to the
Soldi et al.'s values of 400 $\times 10^{22}$ atoms cm$^{-2}$ and
50 keV, respectively. The fit results in a reduced $\chi^2$ of
0.72 (18 dof, see Figure 3, right panel). The spectral index is
then derived as 1.82$\pm$0.09, consistent with Soldi et al.'s
value of 1.8$^{+0.4}_{-0.5}$.

\begin{table*}[th]
\begin{center}
\caption{Results for the spectral fitting of GX 301-3, the
Circinux Galaxy, and 1H 1254-690. The values without error bars
show the corresponding parameters that can not be constrained from
the data. } \vspace{0.2cm}
\begin{tabular}{lllllll}
\hline
\multicolumn{1}{c}{Source Name}&\multicolumn{1}{c}{$N_H$}&\multicolumn{1}{c}{$\alpha$}&\multicolumn{1}{c}{Ec}&\multicolumn{1}{c}{K}&\multicolumn{1}{c}{reduced $\chi^2$}&\multicolumn{1}{c}{dof}\\
\multicolumn{1}{c}{}&\multicolumn{1}{c}{$\times$10$^{22}$atoms cm$^{-2}$}&\multicolumn{1}{c}{}&\multicolumn{1}{c}{keV}&\multicolumn{1}{c}{$\times$keV$^{-1}$cm$^{-2}$s$^{-1}$}&\multicolumn{1}{c}{}&\multicolumn{1}{c}{}\\

\hline
 GX 301-2                &\ldots &2.1$^{+0.2}_{-0.6}$ &4.61$^{+0.15}_{-0.02}$  & 4$\times$10$^{-4}$&0.9&17        \\
 Circinus Galaxy         &400 (fixed)&1.82$\pm$  0.09 &  50 (fixed) & 0.11$^{+0.04}_{0.03}$    &0.7 &18        \\
 1H 1254-690              &\ldots &1.64$\pm$0.33 & 4.5$^{+1.3}_{-0.9}$  &0.62$\pm$0.20 & 1.2&  133      \\

\hline \hline
\end{tabular}
\end{center}
\label{cantotable}
\end{table*}

%For the spectrum of Circinus Galaxy, the following model was used
%to plot the data:
%f(E)=wabs*cutoffpl
%   =exp(-Nh*cro)*k*E**(-alf)*exp(-E/beta)
%where Nh is the column density in units of atoms/cm^2, cro the
%cross section, alf the power law index, beta the cutoff energy.
%These parameters are taken as Nh=400e^22 atoms/cm^-2
%cro=(701.2+25.2*E)*E**(-3) (at energies > 10 keV, see Marrison ApJ
%270, 119). k=0.109790 photons keV^-1 cm^-2 s^-1 at 1 keV
%alf=1.82373 beta=50 keV
%The plot was E^2*f(E), in units of MeV cm^-2 s^-1.

\subsection{Hard X-rays to MeV spectral constraints}

%As discussed in the introduction, Zhang et al. reported that
%GRO~J1411-64 underwent a flare in 1995, lasting for 3 months (7
%viewing periods of COMPTEL) in the energy range 1-3 MeV. The flare
%spectrum was very steep and the extrapolation to the INTEGRAL band
%requires a rather bright hard X-ray source at energies $\le$ 500
%keV.
The light curve of GRO~J1411-64 is shown in Figure 4, including the
flare period, showing that if there is emission out of the flare
it is likely steady in 0.75-1 MeV band, and requires, in absence
of a peak in the spectral energy distribution, a relatively bright
INTEGRAL source. During its low state, the source was detected at
$\sim 4\sigma$ by COMPTEL in the energies of 0.75-1 MeV. In Figure
5, a map of GRO J1411-64 is presented based on data only from the
low state in this energy band. The corresponding spectrum of the
low state can be represented by a power law shape with spectral
index 2.5$^{+0.6}_{-0.4}$ (see Figure 6).

%%%%% Fig 4 %%%%%%%%%%%%%%%%%%%%%%%%%%%%%%%%%%%%%%%%%%%%
\begin{figure}[t]
\centering
\includegraphics[clip=true,scale=.4]{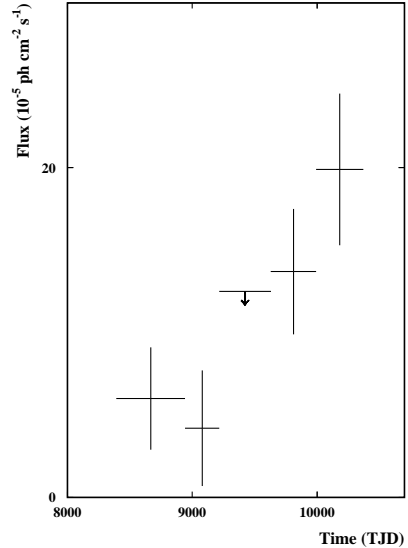}
\caption{Light curve of GRO~J1411-64 as observed by COMPTEL at
0.75-1 MeV band. The error bar is 1$\sigma$ and upper limit 2
$\sigma$. These data points include the 7 viewing periods when the
source was flaring. }
\end{figure}
%%%%% Fig 4 %%%%%%%%%%%%%%%%%%%%%%%%%%%%%%%%%%%%%%%%%%%%

%%%%% Fig 5 %%%%%%%%%%%%%%%%%%%%%%%%%%%%%%%%%%%%%%%%%%%%
\begin{figure}[hb]
\centering
\includegraphics[clip=true,scale=.4]{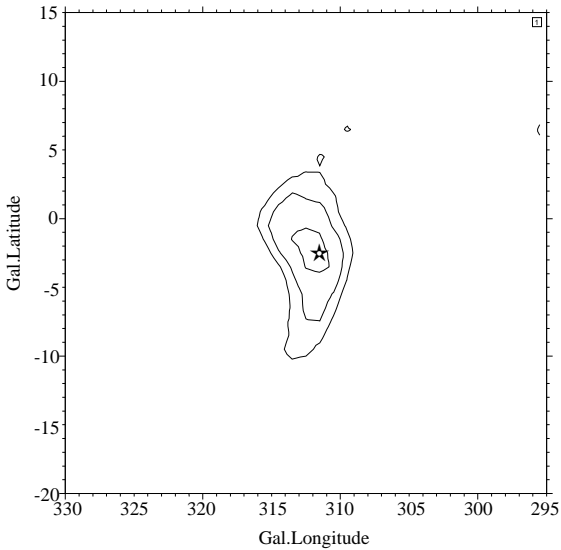}
\caption{The skymap of GRO~J1411-64 as observed by COMPTEL in
0.75-1 MeV during 1991-1996, not including the flare period of 4
months in 1995. The star represents the best-guessed source
location. The contour lines start at a detection significance
level of 3$\sigma$ with steps of 0.5$\sigma$. }
\end{figure}
%%%%% Fig 5 %%%%%%%%%%%%%%%%%%%%%%%%%%%%%%%%%%%%%%%%%%%%

%%%%% Fig 6 %%%%%%%%%%%%%%%%%%%%%%%%%%%%%%%%%%%%%%%%%%%%
\begin{figure}[t]
\centering
\includegraphics[clip=true,height=7cm, scale=.2]{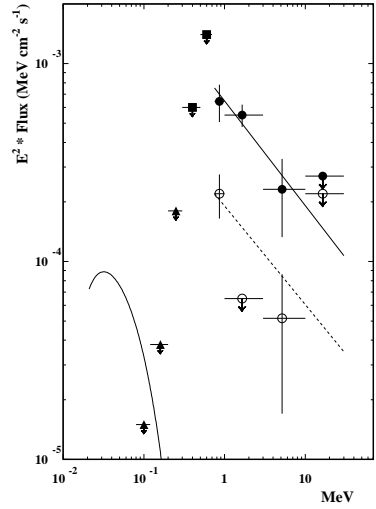}
\caption{Combined energy spectrum of GRO~J1411-64. Filled (open)
circles represent flare (low) states of GRO~J1411-64 as observed
by COMPTEL at MeV energies. The corresponding power law shapes
(solid line for the flare state, dashed line for the low state),
are compared to 2$\sigma$ upper limits obtained from IBIS/ISGRI
(triangles) and SPI (squares). The solid curve at low energies is
the energy spectrum of Circinus Galaxy derived from the IBIS/ISGRI
data. The error bar is 1$\sigma$ and upper limit 2$\sigma$. }
\end{figure}
%%%%% Fig 6 %%%%%%%%%%%%%%%%%%%%%%%%%%%%%%%%%%%%%%%%%%%%

The ISGRI/SPI upper limits combined to spectra of both flare/low
states shows the existence of a maximum in the power output at
hard X-rays. If the source was at its flare state, the power
maximum in the combined spectrum is around 500-700 keV, otherwise,
if the source was at the low state as defined by Zhang et al.
(2002), the maximum is somewhere between 300-700 keV.

In addition, a combination of the best-fit spectrum of the
Circinus Galaxy at hard X-rays with the one detected at MeV
energies of GRO~J1411-64, suggests that this Seyfert galaxy is not
the counterpart of the MeV source: its spectral extrapolation is
well below the ones at MeV energies for both states.  Such a
unique power maximum at high energies, might remind the second
peak in the SED of a blazar or microblazar, which represent the
emission component originating in the inverse Compton process of
the soft photon off the highly relativistic leptons within the
jet. We explore briefly this possibility in the ending discussion.

\section{{\it XMM-Newton} observations of the innermost region of the COMPTEL
source}

The COMPTEL uncertainty in the position of the source, and the
fact that no clear counterpart has been found in the INTEGRAL
range, makes a direct search at keV energies more difficult. In
any case, we have conducted a search of the innermost region of
the COMPTEL location, in order to avoid missing any obvious faint
counterpart that may be residing there, and that might have been
missed by the INTEGRAL observations. The best position of the
COMPTEL source (centered at $(l,b)$ =311.5$^\circ$, -2.5$^\circ$)
had not been observed by any recent X-ray mission (Chandra, ASCA,
XTE, or {\it XMM-Newton}) until now. HEASARC archival search
showed no matches for past X-ray missions (Einstein, Ginga,
Copernicus, etc.) either. { In addition, there is no ROSAT pointed
observation. ROSAT only reported three faint sources about 13 to
20 arcmin from the center of the COMPTEL source, in the ROSAT
Faint Sources Catalog (rassfsc), with only $\sim 400$ s exposure.}

The best localization of COMPTEL source GRO 1411-64 position was
observed with XMM-Newton during revolution 960 on the 7th of March
of 2005 (Observation ID: 0204010101). The EPIC instrument cameras,
MOS1, MOS2 and pn, were operated in full window (imaging) mode,
with thick filters to prevent optical contamination from bright
stars in the field. The exposure times for each camera were 18, 18
and 22 ks for MOS1, MOS2 and pn respectively.
The data were pipeline-processed with the XMM-Newton Science
Analysis Software (SAS) version 6.1.  After removal of background
flares, a total of 15.8, 15.8 and 14.6 ks of good data survived
for MOS1, MOS2 and pn respectively.

\begin{figure}[t]
\centering
\includegraphics[clip=true,scale=.4]{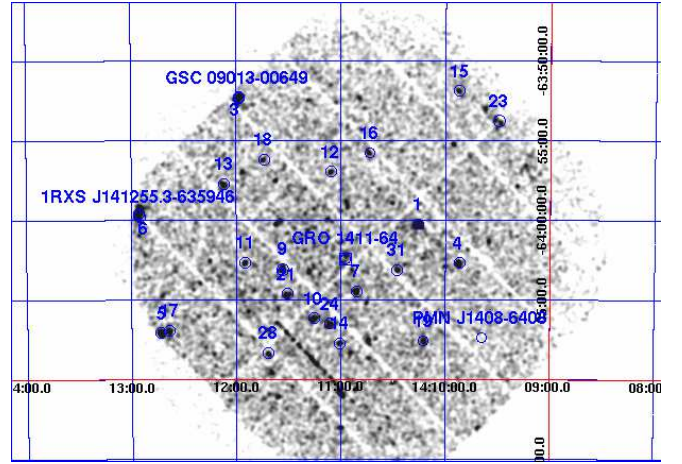}
\caption{XMM-Newton EPIC image (combining all 3 cameras), where
detected sources and previously catalogued sources have been
labeled. The centroid of the COMPTEL source GRO 1411-64 is marked
with a square box, the error contour being larger than the image
itself.} \label{XMMimage}
\end{figure}

A total of 31 X-ray sources were formally detected by the SAS
source detection algorithm in the EPIC data. Nine of these were
excluded due to detector defects and other artifacts, in a careful
inspection. The resulting 22 reliable and statistically
significant sources (likelihood $> 10$) are shown in
Figure~\ref{XMMimage}, and their main properties catalogued in
Table~\ref{XMMtable}. Due to the low Galactic latitude of this
observation, no attempt was made to refine the astrometry provided
by the spacecraft attitude solution against optical sources. This
is known to lead to systematic uncertainties in the position of
the sources of up to a few arcsec at most.

X-ray spectra of all these sources were extracted from the 3 EPIC
cameras, using circular regions with radii that optimize the
signal to noise in every detector. Background spectra were
extracted in nearby circular source-free regions that avoid CCD
gaps. Calibration files (redistribution RMF and effective area
ARF) were generated for every spectrum, using the SAS tasks {\tt
rmfgen} and {\tt arfgen}.

\begin{figure}[hb]
\centering
\includegraphics[clip=true,scale=.25,angle=-90]{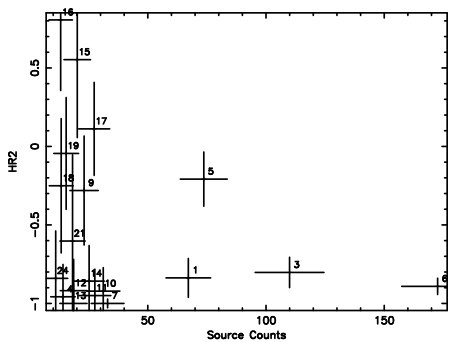}
\caption{EPIC pn hardness ratio $HR_2$ (see text for details)
versus source counts in that detector.} \label{XMMhr}
\end{figure}

In Figure~\ref{XMMhr} we display the X-ray hardness ratio
($HR_2=(H-S)/(H+S)$, where $H$ and $S$ are the flux in the 2.0-4.5
keV and 0.5-2.0 keV bands respectively) versus the total EPIC-pn
source counts of the sources detected in the EPIC-pn camera. Given
the large Galactic column along this direction ($N_H\sim
1.13\times 10^{22}\, {\rm cm}^{-2}$), all extragalactic sources
are expected to have a high value of the hardness ratio $HR_2>0$.
Extragalactic sources are therefore a sizeable fraction of the
sources detected.

\begin{table*}[t]
\begin{center}
\caption{Catalogue of the sources detected in the {\it XMM-Newton}
EPIC-pn camera} \vspace{0.2cm}
\begin{tabular}{ccccccccc}
\hline
\multicolumn{2}{c}{Source}&\multicolumn{3}{c}{Position}&\multicolumn{1}{c}{EPIC-pn 0.2-12 keV}&\multicolumn{1}{c}{EPIC-pn $HR_2$}\\
\multicolumn{1}{c}{Name}&\multicolumn{1}{c}{Label}&\multicolumn{1}{c}{RA(J2000)}&\multicolumn{1}{c}{DEC(J2000)}&\multicolumn{1}{c}{Error
(arcsec)}&\multicolumn{1}{c}{Source
Counts}&\multicolumn{1}{c}{}\\
\hline \hline
XMMU J141015.5-640015 &  1   &14:10:15.5&-64:00:15.7&0.4   & 6$7   \pm 9$  & $-0.84 \pm 0.12$\\
XMMU J141157.9-635216 &  3   &14:11:57.9&-63:52:16.8&0.4   & $110  \pm 15$ & $-0.80 \pm 0.10$\\
XMMU J140952.3-640243 &  4   &14:09:52.3&-64:02:43.2&0.7   & $14   \pm 5$  & $-0.96 \pm 0.21$\\
XMMU J141243.2-640701 &  5   &14:12:43.2&-64:07:01.4&0.6   & $74   \pm 10$ & $-0.21 \pm 0.17$\\
XMMU J141255.6-635929 &  6   &14:12:55.6&-63:59:29.8&0.3   & $173  \pm 15$ & $-0.89 \pm 0.05$\\
XMMU J141051.2-640429 &  7   &14:10:51.2&-64:04:29.2&0.7   & $33   \pm 7$  & $-1.00 \pm 0.03$\\
XMMU J141133.3-640304 &  9   &14:11:33.3&-64:03:04.3&0.9   & $23   \pm 6$  & $-0.28 \pm 0.35$\\
XMMU J141115.6-640610 &  10  &14:11:15.6&-64:06:10.3&1.2   & $31   \pm 7$  & $-0.92 \pm 0.15$\\
XMMU J141154.8-640238 &  11  &14:11:54.8&-64:02:38.5&1.0   & $28   \pm 7$  & $-0.95 \pm 0.17$\\
XMMU J141105.6-635659 &  12  &14:11:05.6&-63:56:59.2&1.7   & $19   \pm 6$  & $-0.92 \pm 0.20$\\
XMMU J141206.6-635743 &  13  &14:12:06.6&-63:57:43.2&1.0   & $19   \pm 6$  & $-1.00 \pm 0.13$\\
XMMU J141100.7-640744 &  14  &14:11:00.7&-64:07:44.7&1.5   & $25   \pm 6$  & $-0.86 \pm 0.23$\\
XMMU J140951.9-635156 &  15  &14:09:51.9&-63:51:56.9&1.3   & $20   \pm 6$  & $+0.55 \pm 0.50$\\
XMMU J141043.3-635547 &  16  &14:10:43.3&-63:55:47.6&1.4   & $13   \pm 5$  & $+0.81 \pm 0.50$\\
XMMU J141238.5-640652 &  17  &14:12:38.5&-64:06:52.0&1.2   & $27   \pm 7$  & $+0.11 \pm 0.30$\\
XMMU J141143.8-635610 &  18  &14:11:43.8&-63:56:10.4&1.2   & $13   \pm 5$  & $-0.25 \pm 0.43$\\
XMMU J141012.6-640736 &  19  &14:10:12.6&-64:07:36.6&1.5   & $15   \pm 5$  & $-0.04 \pm 0.36$\\
XMMU J141130.8-640439 &  21  &14:11:30.8&-64:04:39.8&1.6   & $18   \pm 5$  & $-0.60 \pm 0.55$\\
XMMU J141106.3-640632 &  24  &14:11:06.3&-64:06:32.1&1.1   & $11   \pm 5$  & $-0.84 \pm 0.30$\\
\hline \hline
\end{tabular}
\end{center}
\label{XMMtable}
\end{table*}

We now concentrate on a careful analysis of the spectra of the 4
likely Galactic sources (small hardness ratio) with more than 50
EPIC-pn source counts.

Source XMMU J141015.5-640015 (label 1) is well fitted by a simple
black body spectrum, with no absorption.  The best fit for the
temperature is $kT\sim 270_{-30}^{+40}\, {\rm eV}$, which points
towards a nearby active stellar corona.

Source XMMU J141157.9-635216 (label 3) matches the position of the
star GSC 09013-00649. The X-ray spectrum is well fit by a black
body at $kT\sim 250_{-70}^{+40}\, {\rm eV}$, modified by a small
amount of photoelectric absorption ($N_H\sim
0.15_{-0.14}^{+0.3}\times 10^{22}\, {\rm cm}^{-2}$), and is
therefore consistent with a Galactic star with some intervening
absorption.

Source XMMU J141243.2-640701 (label 5), the hardest of the spectra
analyzed, is most likely an extragalactic AGN, as the X-ray
spectrum is appropriately fit by a power law with energy spectral
index $\Gamma\sim 1.6_{-1.1}^{+0.8}$ and absorbing column
$N_H=0.7_{-0.7}^{+0.6}\times 10^{22}\, {\rm cm}^{-2}$, consistent
with the Galactic column in that direction.

Source XMMU J141255.6-635929 (label 6, coincident with
$ROSAT$source 1RXS J141255.3-635946 within 16 arcsec) is very
intriguing. In the discussion that follows, we consider only the
energy range from 0.2-7 keV, for which we have 51 spectral bins
with 10 counts each. The source has a small hardness ratio,
indicating a Galactic origin. However, its spectrum is not well
fit by an absorbed thermal model ($\chi^2=75.8$, for 48 degrees of
freedom) which leaves large and correlated residuals both at low
and high energies. The low energy residuals can be removed if we
restrict the fit to below 2 keV, but then a significant high
energy tail is apparent. The origin of this high energy tail can
be addressed in at least two ways. First, we can fit a second
black body spectrum with a different absorption. This gives a
formally acceptable fit to the whole spectrum ($\chi^2=49.70$ for
46 degrees of freedom), with a low-temperature black body with
$kT=121_{-39}^{+30}\, {\rm eV}$ absorbed by
$N_H=0.39_{-0.17}^{+0.29}\,\times 10^{22}\, {\rm cm}^{-2}$ and a
second much hotter black body with $kT=432_{-120}^{+404}\, {\rm
eV}$ absorbed by a highly uncertain column $N_H=2_{-2}^{+6}\,
\times 10^{22}\,{\rm cm}^{-2}$ (as customary, errors denote 90\%
uncertainty for one single interesting parameter). This second
black body appears very large for a second active star,
positionally confused with the former. Although we cannot reject
this option, we find it very unlikely. An additional possibility
is that the hard excess in this source is modeled as a power law.
The fit to a black body plus a power law, both absorbed by
intervening material, is also very good with $\chi^2=51.90$ for 46
degrees of freedom.  In this case $kT=106_{-30}^{+23}$,
$N_H=0.48_{-0.18}^{+0.42} \times 10^{22}$ cm$^{-2}$ and a highly
uncertain photon index of $\Gamma=2.3_{-0.8}^{+1.2}$. Attempts to
fit the soft part of the spectrum with a steep power law resulted
in extremely large values of the spectral slope ($\sim 7-8$) and
significant residuals accompanied by a large increase in the
$\chi^2$.

\begin{figure}[hb]
\centering
\includegraphics[clip=true,scale=.3,angle=270]{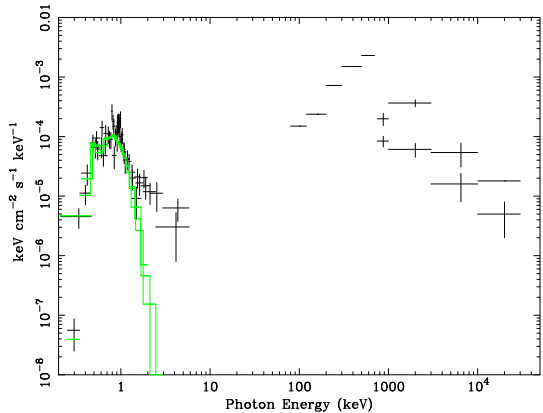}
\caption{XMM-Newton unfolded spectrum of the X-ray source XMMU
J141255.6-635932. The model shown is only the thermal component in
the X-ray spectrum. The COMPTEL detections and the INTEGRAL upper
limits are also shown at high energies, with horizontal bars
denoting 2$\sigma$ upper limits.} \label{XMMspec}
\end{figure}

In Figure~\ref{XMMspec} we plot the unfolded spectrum (largely
independent of the model fitted) for XMMU J141255.6-635932, along
with the best fit model (with these parameters), the COMPTEL
detections and the INTEGRAL upper limits from Table 1. The hard
excess exhibited by the XMM-Newton data is apparent in that
figure, and might be suggestive of a large Compton bump that would
peak in the several $\sim 100$ keV region, fitting well with the
COMPTEL detections. However, the fact that the XMM-Newton image
does not cover the full COMPTEL source location and the
non-detection by INTEGRAL of any reliable counterpart, would make
the assumption that XMMU J141255.6-635932 is the counterpart to
GRO 1411-64, although spectrally consistent, only tentative and
risky.

\section{Discussion and concluding remarks}

\begin{figure}[hb]
\centering
\includegraphics[clip=true,scale=.4]{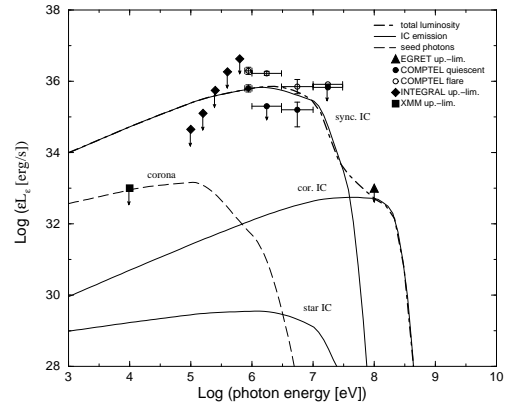}
\caption{ A microquasar model in the light of observational
constraints. See text for details. } \label{hecomptel}
\end{figure}

We have applied to the non-simultaneous data from XMM-Newton
(treating them as an upper limit\footnote{This is actually a not
well-justified a priori assumption, since the coverage of
XMM-Newton observations is much smaller than that of INTEGRAL, and
it is imposed by fiat, in the knowledge that the constraints
imposed below do not crucially depend on the XMM-Newton point, but
rather on the steepness of the INTEGRAL upper limits.}), INTEGRAL
and COMPTEL the microquasar model presented by Bosch-Ramon et al.
(2006). This model consists of a magnetized conical jet filled
with relativistic electrons which radiate through synchrotron and
inverse Compton scattering with star, disk, corona and synchrotron
photons. These electrons lose energy mainly through adiabatical
losses, as a first order approximation.
At the top of Table~\ref{common}, parameter values for a typical
microquasar system and jet geometry are given (Bosch-Ramon et al.
2006). We have considered different values within the range open
for the free parameters finding that it is not possible to obtain
a simple microquasar model that could fit the SED. In particular,
in Figure~\ref{hecomptel} we show a test case with the free
parameters fixed to the values presented in Table~\ref{common}, at
the bottom. { We note that, since the computed SED in Fig. 10 is
dominated in the gamma-ray band by SSC emission, the model results
would also apply for a low mass microquasar.}

\begin{table*}[]
  \caption[]{Parameter values for GRO~1411$-$64}
  \label{common}
  \begin{center}
  \begin{tabular}{cl}
  \hline\noalign{\smallskip}
Parameter &  values \\
  \hline\noalign{\smallskip}
Stellar bolometric luminosity [erg~s$^{-1}$] & $10^{38}$ \\
Distance from the apex of the jet to the compact object [cm] & $5\times10^7$ \\
Initial jet radius [cm] & $5\times10^6$ \\
Orbital radius [cm] & $3\times10^{12}$  \\
Viewing angle to the axis of the jet [$^{\circ}$] & $45$ \\
Jet Lorentz factor & 1.2 \\
\end{tabular}
\begin{tabular}{cl}
\hline\noalign{\smallskip} \hline\noalign{\smallskip}
Jet leptonic kinetic luminosity [erg~s$^{-1}$] & $3\times10^{35}$ \\
Maximum electron Lorentz factor (jet frame) & 5$\times10^2$ \\
Maximum magnetic field [G] & 8000 \\
Electron power-law index & 1.5 \\
Total corona luminosity [erg~s$^{-1}$] & $3\times10^{33}$ \\
  \noalign{\smallskip}\hline
  \end{tabular}
  \end{center}
\end{table*}

%From the figure, we see that XMM observations, if average values
%are assumed for the whole region, would imply quite low
%luminosities (up to 10$^{33}$~erg~s$^{-1}$) for the X-ray sources.
%The upper-limits obtained with INTEGRAL are not so restrictive,
%although the upper-limits around 100~keV are well below the
%calculated luminosity. Finally, the COMPTEL fluxes, when out of
%flare period, are not too far below the flare fluxes, although
%there are upper-limits at energies between those at which the
%source was detected. All this implies that

If the COMPTEL source is to be among the group of low luminosity
soft X-ray sources and be detectable at energies about 1~MeV, a
very hard photon index of about 0.67 (i.e. $n_{\nu}\propto
\nu^{-0.67}$) will be required in the  energy range 10--1000~keV
assuming a power-law energy distribution for the radiation. Even
for a strongly absorbed soft X-ray counterpart (see, e.g. Walter
et al. 2003), or a soft X-ray counterpart out of the XMM-Newton
field observed, the INTEGRAL upper-limits impose also a very hard
photon index of about 1 in the range 100-1000~keV, whereas
radiation at higher energies is strongly constrained by EGRET
observations. In the context of microquasars, if the particle
spectrum follows a power-law, it is difficult to explain such
narrow band emission concentrated in the MeV range.

{ Concerning other transients spatially coincident with the
INTEGRAL observations we mention 2S 1417-624, a Be/X-ray binary
pulsar system, with a Be star of $>5.9$ M$_\odot$ for a neutron
star of 1.4 M$_\odot$. BATSE observed a huge outburst of this
source at hard X-rays (pulsations up to 100 keV were detected),
which started on August 26, 1994 and lasted 110 days (Finger et
al., 1996). COMPTEL observed the source twice (VPs 402, 402.5)
during this period but did not detect it. During the next 200
days, 2S 1417-624 showed 5 smaller outbursts (Finger et al.,
1996), which coincide in time with the significant COMPTEL
detection of the MeV source in 1995. During the huge outburst in
hard X-rays, the pulsed luminosity was found to be much less than
it would be estimated from the spin-up rate. This implies that
most of the power output is either unpulsed or outside of the hard
X-ray range (Finger et al., 1996). Assuming 2S 1417-624 as
counterpart of the COMPTEL source, its 0.75-10 MeV luminosity
would be $2.5 \times 10^{37}$ erg s$^{-1}$, obtained by using its
upper-limit distance of 11.1 kpc. This luminosity is comparable to
the pulsed one of $2.2 \times 10^{37}$ erg s$^{-1}$. 2S 1417$-$624
as counterpart might show an anticorrelation of the X- and
$\gamma$-ray emission as proposed for other sources (Romero et al.
2001). GS 1354-645, also in the field, is a transient black-hole
X-ray binary (XRB) system. However, no X-ray observations of GS
1354-624 during the activity period reported for the COMPTEL
source are reported. Since INTEGRAL did not detect these sources
during our campaign, we cannot open judgment about the possibility
for an outbursting system such as these to be the counterpart of
the reported detection. }

% re-look at Comptel source
%
Failing to associate a plausible counterpart of GRO~J1411-64 at
the hard X-rays from our INTEGRAL observations, we have looked
again at the initially reported gamma-ray properties of
GRO~J1411-64. In Table 1, we have already expanded beyond the
initially reported source characteristics from Zhang et al. (2002)
by including the individual detection significances. If we apply a
rather conservative detection criterion (significance threshold)
of at least 4 $\sigma$ in an individual measurement, a firm
detection can be claimed only in the so-called flare state, which
was attributed to occur within CGRO viewing periods 414-424.
During this flare period as well as any other investigated time
span, a detection threshold of at least 4$\sigma$ in the
individual measurement allows the determination of the photon
spectrum from significantly measured data points, i.e., in two
energy bands only, from 0.75-1 MeV and 1-3 MeV, respectively. An
accordingly determined spectrum yields an index of 2.18 $\pm$ 0.2,
substantially harder than the previous reported index of 2.53
$\pm$ 0.2 taking into account lower confidence data points. If we
fit the range from 0.75 to 10 MeV with a statistically weighted
power law fit, an index of 2.48 $\pm$ 0.2 will be determined.
Therefore, the dependence from the individual measurements becomes
crucial for determining a spectral slope.

\begin{figure}[hb]
\centering
\includegraphics[clip=true,scale=1.]{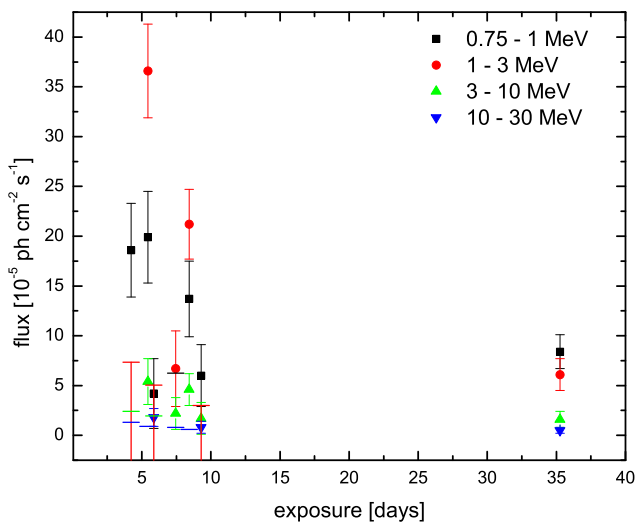}
\caption{The original observations of GRO~J1411-64 as reported in
Zhang et al. (2002), see Table 1. The dependence from the
individual measurements from the underlaying exposure is easily
seen, as is the different character in the phase 4 vs phase 5
measurements, which exhibit the largest flux values, but missing a
1-3 MeV flux in case of the phase 5 measurement.} \label{XXX}
\label{exposurevsflux}
\end{figure}

If we compare the observations of GRO~J1411-64, as made in Figure
\ref{XXX}, indicating the measured flux/upper limit in the four
energy bands as a function from the respective exposure, other
characteristics can be seen. Here, we convert the 2$\sigma$ upper
limits from Zhang et al. (2002) into 1$\sigma$ upper limits, in
order to prevent quoting upper limits based on higher statistical
confidence than certain flux values from Table 1. One clearly sees
that the flaring phenomenon related to the CGRO observation
periods 414-424, is not based on any extraordinary long or short
exposure when compared with other observations. It can also be
seen that the best signals originated from the two lower energetic
band in COMPTEL, as already indicated from the individual
detection significances. Finally, whereas there is a clear
indication of the spectral behavior being a steep spectrum source
in the phase 4 observations, there is a remarkable difference in
the phase 5 observation: Here, the largest flux has been quoted
from the 0.75 to 1 MeV band. However, in the immediate neighboring
energy band (1 -3 MeV), only an upper limit has been obtained.
Comparing for the so-called flare phase, where the flux in the
0.75-1 MeV band is clearly exceeded by the 1-3 MeV band data, we
should perhaps reconsider the flare phenomenon: On one hand, the
actual low energy band data point in phase 5 is measured at higher
flux and better statistical significance, on the other hand the 1
-3 MeV flux measurement is missing, indicating that the spectrum
was even more extreme in the phase 5, and clearly different from
phase 4, and any other considered time span quoted in Zhang et al.
(2002). This is also seen in the total superposition, where the
four energy bands line up in an more usual behavior, i.e., any
subsequent higher energetic band exhibit a lower flux. In the
light of such reassessment we note that a flare phenomenon is not
consistently to be described as either spectral hardening or
softening compared to the sparse overall data from the region. We
note, however, that the indication for this phenomenon is
primarily based on data with the lowest effective exposure. Thus,
the distinction of different state becomes blurry.

%Perhaps, a better
%approach would be a compact jet source without significant X-ray
%emission produced by accretion, as in the model presented by
%Punsly et al. (2000).
%, where an electron-positron beam is produced
%by magnetic effects associated with an isolated Kerr-Newman black
%hole. In any case,

GLAST observations would help improving the location of the MeV
source if radiation at higher energies is not completely
suppressed, and would open the door for more efficient
multiwavelength searches of the counterpart. However, it is true
that the nature of this COMPTEL source might not be constrained
further if this detection was a one-time only transient phenomena.
GLAST will only be able to help if a candidate counterpart is
caught in the act (flaring/quiescent state of an AGN or a more
rare galactic object). Having at hand GLAST observations, in any
case, will make our currently reported investigation to naturally
fit into the testing of any hypothesis on the nature of GRO
J1614-64.

 \acknowledgements

We thank Dr. M.T. Ceballos for her help with the XMM-Newton data.
DFT was partially supported by NASA and acknowledges the Lawrence
Livemore National Laboratory, where the initial part of this
research was done. S. Zhang was subsidized by the Special Funds
for Major State Basic Research Projects and by the National
Natural Science Foundation of China. XB and AC were financially
supported for this research by the Ministerio de Educaci\'on y
Ciencia (Spain), under project ESP2003-00812. VB-R and JMP have
been supported by Ministerio de Educaci\'on y Ciencia (Spain)
under grant AYA-2004-07171-C02-01, as well as additional support
from the European Regional Development Fund (ERDF/FEDER). VB-R has
been additionally supported by the DGI of the Ministerio de
(Spain) under the fellowship BES-2002-2699. GER was supported by
grants PIP 5375 y PICT 03-13291.

\end{document}